\documentclass[twocolumn,nopacs,floatfix,preprintnumbers,nofootinbib,superscriptaddress]{revtex4}
\usepackage[utf8]{inputenc}
\usepackage[sort&compress]{natbib}
\usepackage{ulem}
\usepackage{bm}
\usepackage{times}
\usepackage{amssymb,amsbsy,amsmath,amsfonts}
\usepackage{graphicx} 
\usepackage{float}
\usepackage{color}
\usepackage{morefloats}
\usepackage{rotating}
\usepackage{srcltx}
\usepackage{overpic}
\usepackage{slashed}
\usepackage{subfigure}
\usepackage{multirow}
\usepackage{verbatim}
\usepackage{hyperref}
\usepackage{tabularx}

\begin{document}

\title{Likely existence of bound states and the Efimov effect in the triple-$J/\psi$ system}

\author{Ya-Wen Pan}
\affiliation{School of Physics, Beihang University, Beijing 102206, China}
\affiliation{Research Center for Nuclear Physics (RCNP), Ibaraki, Osaka 567-0047, Japan}

\author{Zhi-Wei Liu}
\affiliation{School of Physics, Beihang University, Beijing 102206, China}

\author{Li-Sheng Geng}\email{lisheng.geng@buaa.edu.cn}
\affiliation{School of Physics, Beihang University, Beijing 102206, China}
\affiliation{Beijing Key Laboratory of Advanced Nuclear Materials and Physics, Beihang University, Beijing 102206, China}
\affiliation{Peng Huanwu Collaborative Center for Research and Education, Beihang University, Beijing 100191, China}
\affiliation{Southern Center for Nuclear-Science Theory (SCNT), Institute of Modern Physics, Chinese Academy of Sciences, Huizhou 516000, China}

\author{Atsushi Hosaka}\email{hosaka@rcnp.osaka-u.ac.jp}
\affiliation{Research Center for Nuclear Physics (RCNP), Ibaraki, Osaka 567-0047, Japan}
\affiliation{Advanced Science Research Center, Japan Atomic Energy Agency, Tokai, Ibaraki 319-1195, Japan}

\author{Xiang Liu}\email{xiangliu@lzu.edu.cn}
\affiliation{School of Physical Science and Technology, Lanzhou University, Lanzhou 730000, China}
\affiliation{Research Center for Hadron and CSR Physics, Lanzhou University and Institute of Modern Physics of CAS, Lanzhou 730000, China}
\affiliation{Lanzhou Center for Theoretical Physics, Key Laboratory of Quantum Theory and Applications of MoE, Key Laboratory of Theoretical Physics of Gansu Province, Lanzhou University, Lanzhou 730000, China}
\affiliation{MoE Frontiers Science Center for Rare Isotopes, Lanzhou University, Lanzhou 730000, China}

\begin{abstract}

The ground-breaking discovery of the first fully charmed tetraquark state $X(6900)$ in the $J/\psi J/\psi$ invariant mass distribution by the LHCb collaboration has inspired intensive theoretical studies. Various interpretations, such as molecular states, compact tetraquark states, and coupled-channel effects, have been proposed for these states. Of particular interest is the ongoing search for the triple-$J/\psi$ state--a fully-charmed hexaquark state. To deepen our understanding of the triple-$J/\psi$ state and to guide future experimental searches, we study the triple-$J/\psi$ system in this work employing the Gaussian expansion method and the $J/\psi J/\psi$ potential parameterized to yield a shallow bound state, as suggested in several theoretical works. Our results support a triple-$J/\psi$ bound state, even in cases where the attractive interaction between the two $J/\psi$ mesons is very weak. Moreover, our analysis 
implies the Efimov effect in the triple-$J/\psi$ system. In addition, we extend our investigation to the triple-$\Upsilon (1S)$ system and obtain results similar to those for the triple-$J/\psi$ system.

\end{abstract}

\maketitle

\section{Introduction}\label{Intro}

In 2020, the LHCb Collaboration observed the first fully charmed tetraquark state $X(6900)$  in the $J/\psi J/\psi$ invariant mass distribution in $pp$ collisions with a global significance larger than five standard deviations~\cite{LHCb:2020bwg}. 
The CMS Collaboration later confirmed such a state with a local statistical significance of 9.8 standard deviations and reported two new structures, $X(6600)$ and $X(7100)$~\cite{Zhang:2022toq, CMS:2023owd}.
The ATLAS collaboration also observed the $X(6900)$ state with a statistically significant excess over the backgrounds in both the $J/\psi J/\psi$ and $J/\psi\psi'$ invariant mass distributions~\cite{ATLAS:2023bft}.

These discoveries have inspired many theoretical studies of the fully-charmed tetraquark states~\cite{Weng:2020jao, Ling:2021bir, Lu:2020cns, Bedolla:2019zwg, Zhuang:2021pci, Wu:2024euj, Faustov:2022mvs, Zhang:2022qtp, An:2022qpt, Zhang:2020hoh, Gong:2020bmg, Dong:2021lkh, Wang:2022jmb} and various explanations for the experimentally observed states have been proposed, such as compact tetraquark states~\cite{Kuang:2023vac, Anwar:2023fbp, Zhou:2022xpd, Agaev:2023gaq, Chen:2022mcr, Agaev:2023rpj}, molecular states~\cite{Lu:2023ccs, Agaev:2023rpj}, and coupled-channel effects~\cite{Liang:2021fzr, Liang:2022rew, Ortega:2023pmr, Niu:2022jqp}. 
Ref.~\cite{Zhou:2022xpd} argued that $X(6900)$ should be a compact tetraquark state with preferred quantum numbers $0^{++}$ with the pole counting rule.
While Ref.~\cite{Lu:2023ccs} suggested that $X(6900)$ may be a molecular state of higher states, for example, $J/\psi \psi(3770)$ or $\chi_{c0}\chi_{c2}$, based on the standard pole counting rule analysis.  
In Ref.~\cite{Gong:2020bmg}, the authors argued that the Pomeron exchange mechanism could naturally explain the nontrivial structures in the di-$J/\psi$ spectrum observed by the LHCb collaboration. With the coupled-channel interactions between $J/\psi J/\psi$, $J/\psi \psi'$, and $\psi' \psi'$, the narrow structure $X(6900)$ can be explained as a dynamically generated resonance.
In addition to these experimentally observed states, $X(6200)$ has been theoretically predicted as a fully-charmed tetraquark state near the $ J/\psi J/\psi$ threshold.
In Refs.~\cite{Wang:2020wrp,Dong:2020nwy}, the authors described the LHCb $J/\psi J/\psi$ invariant mass spectrum in a coupled-channel approach and found a bound or virtual state with a binding energy of a few MeV near the $J/\psi J/\psi$ threshold called $X(6200)$~\cite{Dong:2020nwy}. With the further accumulation of the di-$J/\psi$ invariant mass spectrum data from the CMS Collaboration, the authors of Ref.~\cite{Wang:2022jmb} revisited this issue
and found that the coupled-channel approach still plays a crucial role in explaining these reported enhancements.
Ref.~\cite{Nefediev:2021pww} employed the QCD string approach to evaluate the mass of the lowest fully-charmed tetraquark state and argued that $X(6200)$ is favored to be a shallow bound state with the preferred spin-parity quantum numbers $0^{++}$ considering the interplay of quark and molecular dynamics. 
In Ref.~\cite{Dong:2021lkh}, the authors argued that the correlated $\pi\pi$ and $K\bar{K}$ exchanges can provide considerable attraction to the $J/\psi J/\psi$ system, leading to the existence of a $J/\psi J/\psi$ bound state. 

Due to their unique properties, fully-heavy multiquark states have always been an intriguing research topic. After the ground-breaking discovery of the fully-charmed tetraquark states, the CMS collaboration has observed the simultaneous production of triple-$J/\psi$ mesons in $pp$ collisions~\cite{CMS:2021qsn}. In this work, we investigate the triple-$J/\psi$ system by employing the Gaussian expansion method (GEM)  with the aim of deepening our understanding of the fully-heavy hexaquark system and providing guidance for future experimental searches.

This article is organized as follows. In Sec.~\ref{frame}, we briefly explain how the GEM is used to solve the three-body Schr\"{o}dinger equation and how to construct the interaction between two $J/\psi$ mesons. The results are presented and discussed in Sec.~\ref{re-dis}, followed by a summary in Sec.~\ref{Con}.

\section{Theoretical Framework}
\label{frame}
~

We solve the three-body Schr\"{o}dinger equation with the  Gaussian Expansion Method (GEM), which has been widely applied to study few-body systems~\cite{Hiyama:2003cu, Liu:2024uxn, Pan:2023zkl, Wu:2022ftm, Pan:2022xxz,Wu:2021kbu,Wu:2019vsy}:
\begin{equation}
    [T+\sum_{1=i<j}^{3}V(r_{ij})-E]\Psi_{J}^{Total}=0\,,
\end{equation}
where $T$ is the kinetic energy, $V(r_{ij})$ 
the two-body potentials between particles $i$ and $j$, and $\Psi_{J}^{Total}$ 
the total wave function of the three-body system, which can be expressed as the sum of the wave functions of the three rearrangements of the Jacobi coordinates:
\begin{equation}
    \Psi_{J}^{Total}=\sum_{c,\alpha}C_{c,\alpha}\Phi_{J,\alpha}^{c}(\pmb{r}_{c},\pmb{R}_{c})\quad (c=1-3)\,,
\end{equation}
with $C_{c,\alpha}$ the expansion coefficients. The lower index $c$ denotes the three Jacobi coordinates we call channels. 
Explicitly, for $c=1$, $\pmb{r} = \pmb{r}_1 - \pmb{r}_2$, and 
$\pmb{R} = (\pmb{r}_1 + \pmb{r}_2)/2 - \pmb{r}_3$, and for $c= 2, 3$ their cyclic ones.  
Note that the three channels should be symmetrized according to the Bose-Einstein statistics for the identical particles $J/\psi$. 
The functions $\Phi_{J,\alpha}^{c}(\pmb{r}_{c},\pmb{R}_{c})$ are the wave functions of each Jacobi coordinate, which are given by
\begin{equation}
    \Phi_{J,\alpha}^{c}(\pmb{r}_{c},\pmb{R}_{c})=\left[\phi^{G}_{n_{c}l_{c}}(\pmb{r}_{c})\psi^{G}_{N_{c}L_{c}}(\pmb{R}_{c})\right]_{\Lambda} \,,
\end{equation}
where $\alpha$ is a collection of 
labels $\{l, L, \Lambda\}$ in which $l$ and $L$ are the orbital angular momenta of the coordinates $r$ and $R$ in each channel, and $\Lambda$ is the total orbital angular momentum. In this work, we only study the $S$-wave case. 
The functions $\phi^{G}_{n_{c}l_{c}}$ and $\psi^{G}_{N_{c}L_{c}}$ are the spatial wave functions which can be expanded in terms of the Gaussian functions of the Jacobi coordinates $\pmb{r}$, $\pmb{R}$:
\begin{equation}
    \begin{aligned}
    &\phi^{G}_{nlm}(\pmb{r})=\phi^{G}_{nl}(r)Y_{lm}(\hat{\pmb{r}})\,, \phi^{G}_{nl}(r)=N_{nl} r^{l} e^{-\nu_{n}r^{2}}\,, \\
    &\psi^{G}_{NLM}(\pmb{R})=\psi^{G}_{NL}(r)Y_{LM}(\hat{\pmb{R}})\,, \psi^{G}_{NL}(R)=N_{NL} R^{L} e^{-\lambda_{N}R^{2}}\,,
    \end{aligned}
\end{equation}
where  $N_{nl}$ and $N_{NL}$ are the  normalization constants.
The parameters $\nu_{n}$ and $\lambda_{N}$ are given by
\begin{equation}
   \begin{aligned}
    &\nu_{n}=1/r^{2}_{n},~~ r_{n}=r_{1}a^{n-1},~~~~~~~~~~(n=1, 2, ..., n_{max})\\
    &\lambda_{N}=1/R^{2}_{N},~~ R_{N}=R_{1}A^{N-1},~~(N=1, 2, ..., N_{max})
   \end{aligned}
\end{equation}
where $\{n_{max}, r_{1}, a~ \mbox{or}~r_{max} \}$ and $\{N_{max}, R_{1},  A~ \mbox{or}~ R_{max} \}$ are Gaussian basis parameters. In this work, we set the parameters as follows: $n_{max}=40$, $N_{max}=30$, $r_{min}=R_{min}=0.01$ fm, and $r_{max}=R_{max}=1000$ fm.
These parameters are optimized by trial and error. The chosen parameter sets are sufficient for our purposes, and further optimization will not change the results.

Although there is insufficient experimental information to determine the $J/\psi J/\psi$ interaction, we 
expect that the interaction between two $J/\psi$ mesons is at least attractive~\cite{Nefediev:2021pww, Dong:2021lkh}. Because of the OZI rule, the exchange of light mesons between the two $J/\psi$ mesons is suppressed. Therefore, the virtual gluon exchanges play a dominant role. This suggests that the $J/\psi J/\psi$ interaction is inherently short-ranged. 
Therefore, 
we employ two functional forms, one is the Gaussian and the other the tripole form of a cutoff $R$ (corresponding to the interaction range),
\begin{eqnarray}
    &&V(r)=C\frac{e^{-(r/R)^2}}{\pi^{3/2}R^3}=C' e^{-(r/R)^2}\, ,  \\ 
    &&V'(r)=C\frac{\pi^2}{4R^3}\frac{1}{(r^{2}+R^{2})^{3}}= C'' \frac{1}{(r^{2}+R^{2})^{3}} \, ,
\end{eqnarray}
where 
$C'$ and $C''$ are the coupling strengths of the potential. Since the radius of the $J/\psi$ is about 0.47 fm according to Ref.~\cite{Eichten:1979ms}, 
a reasonable value for the cutoff $R$ is around 0.5 fm. 
We have mentioned in Sec.~\ref{Intro} that a $J/\psi J/\psi$ 
two-body system may form a shallow bound state.
Therefore, the coupling strengths are determined to yield a few MeV binding energies. 
In this respect, the cutoff $R$ and the binding energies are input parameters in the present study.

\begin{figure*}[!htbp]
  \centering
  {\includegraphics[width=0.33\textwidth]{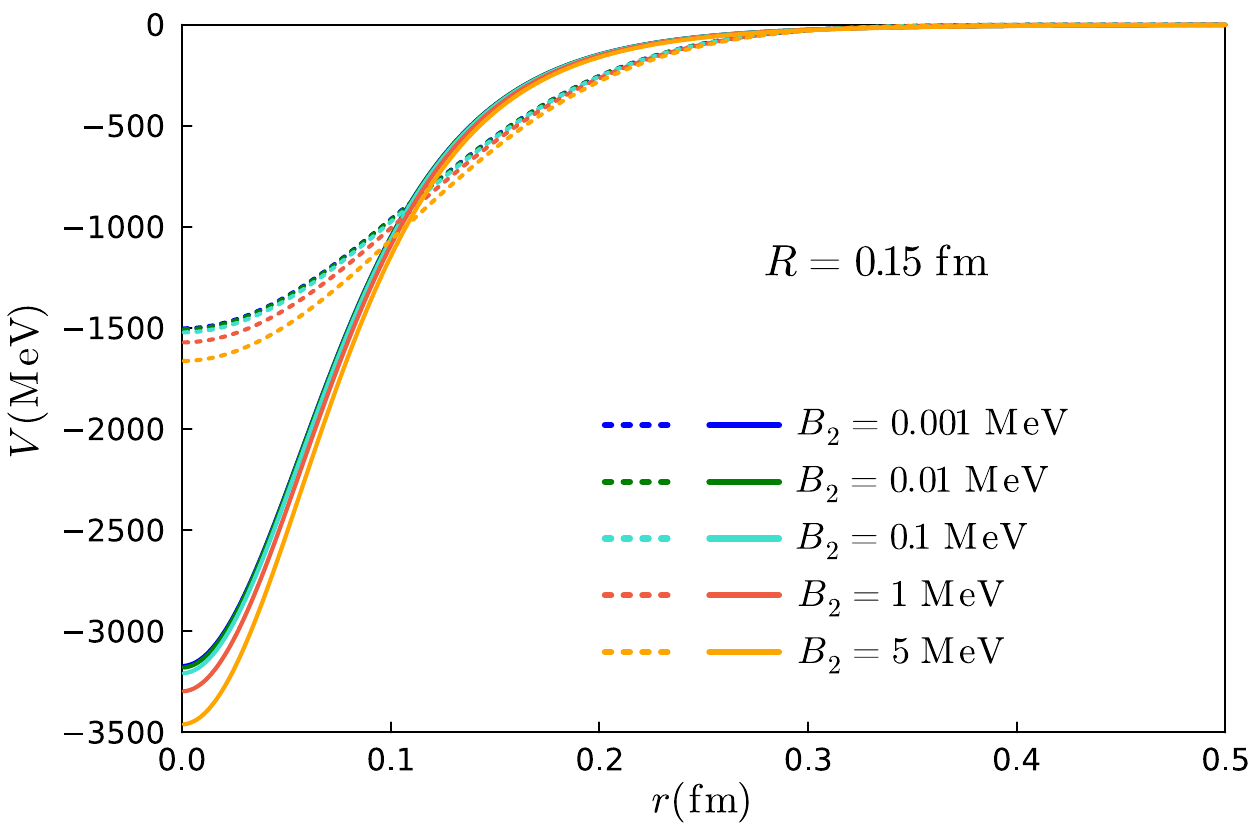}}
  {\includegraphics[width=0.33\textwidth]{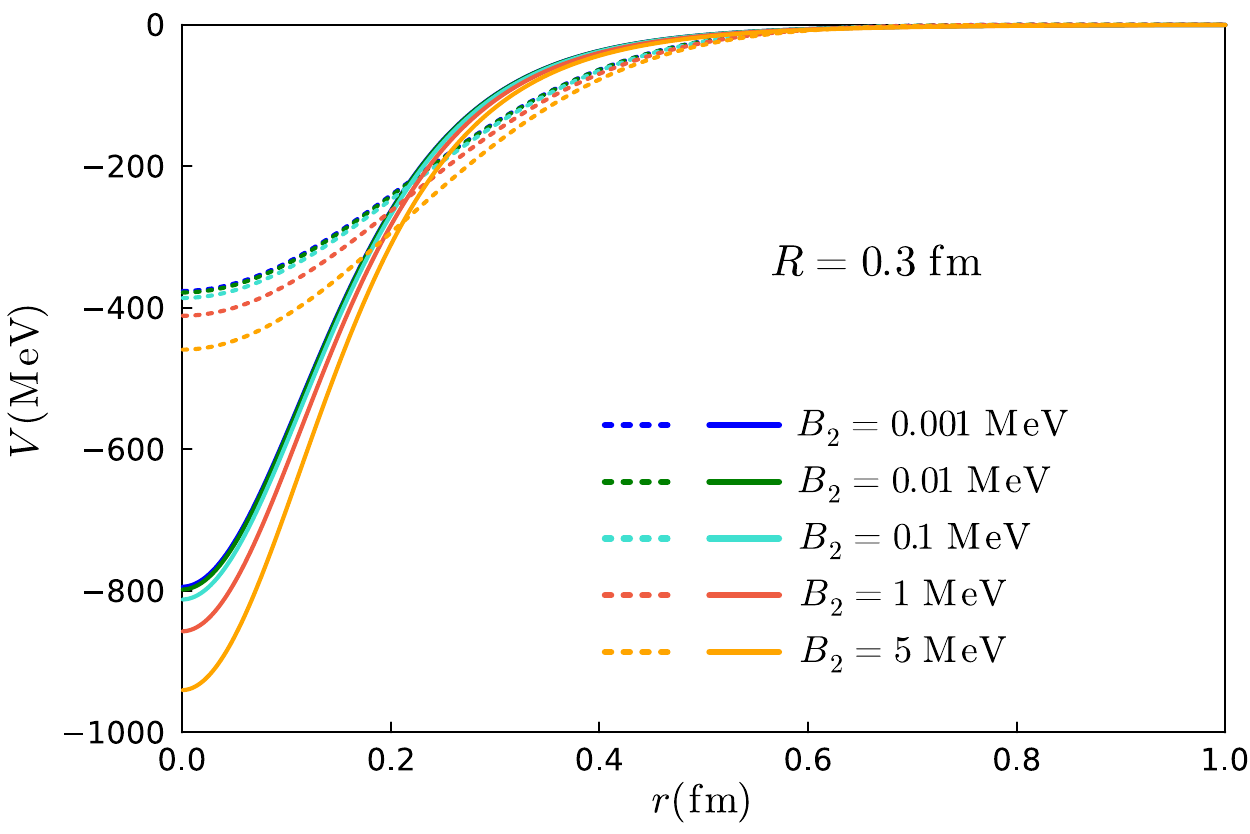}}
  {\includegraphics[width=0.33\textwidth]{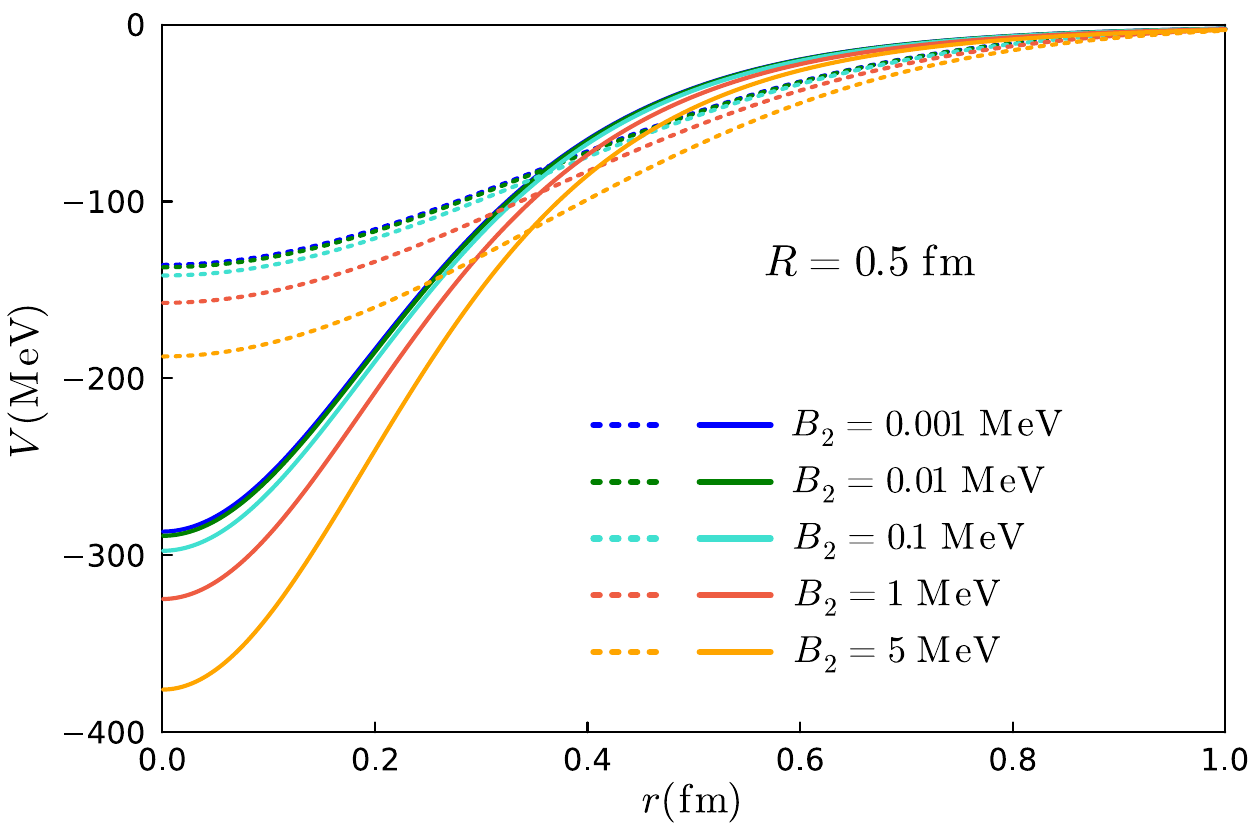}}
  {\includegraphics[width=0.33\textwidth]{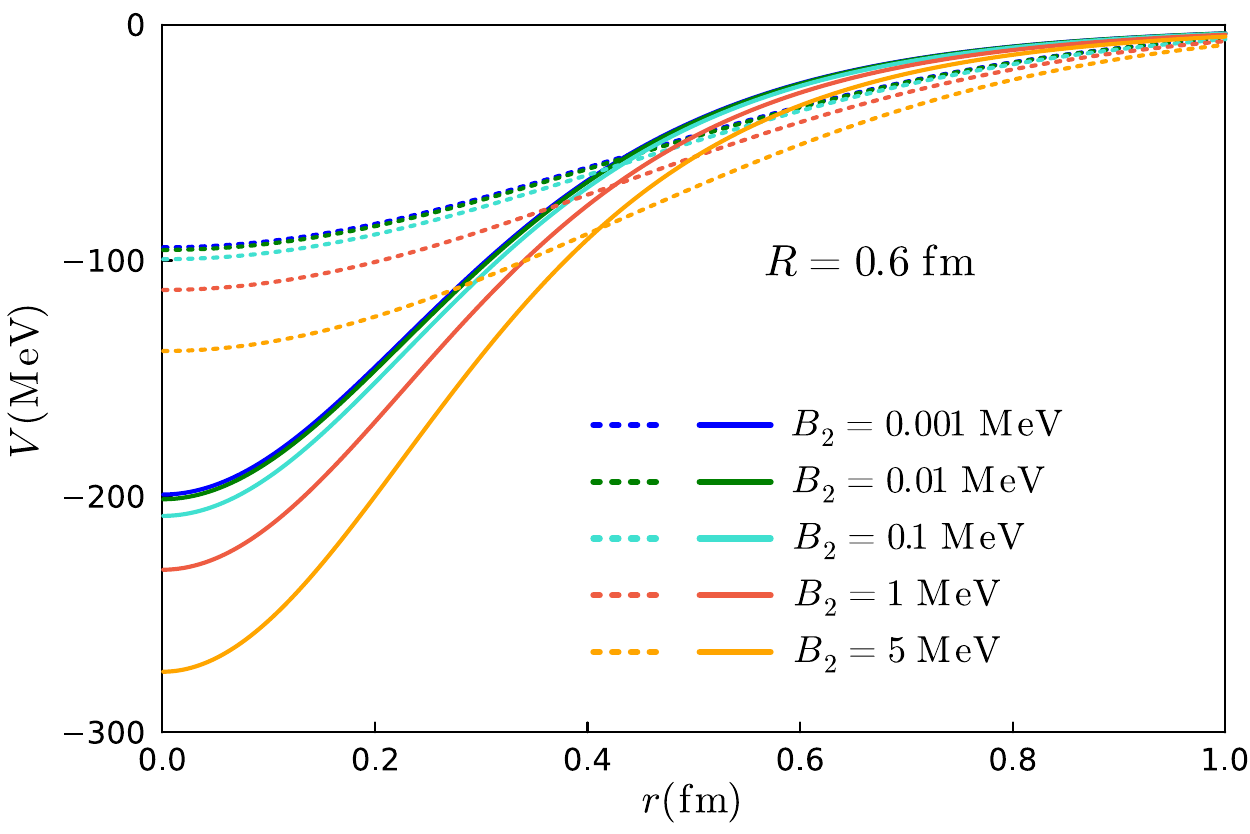}}
  {\includegraphics[width=0.33\textwidth]{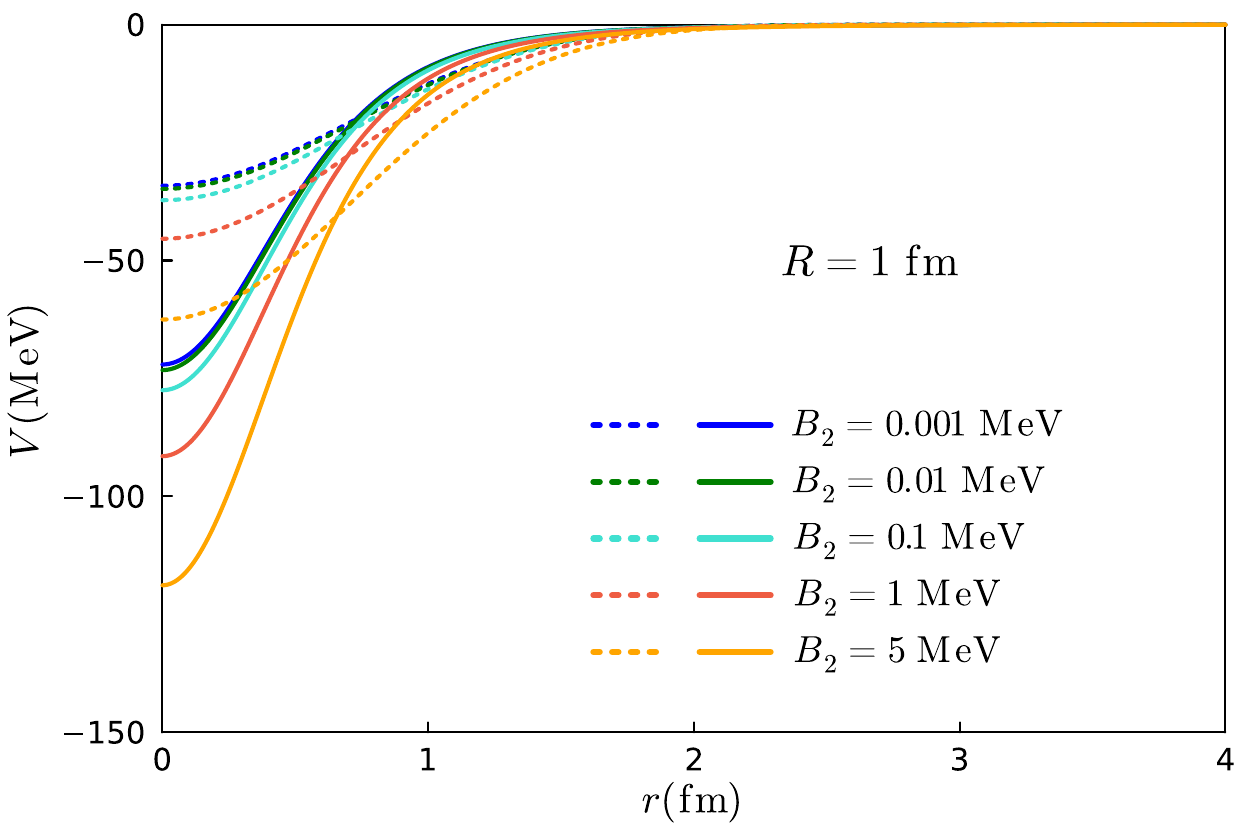}}
  {\includegraphics[width=0.33\textwidth]{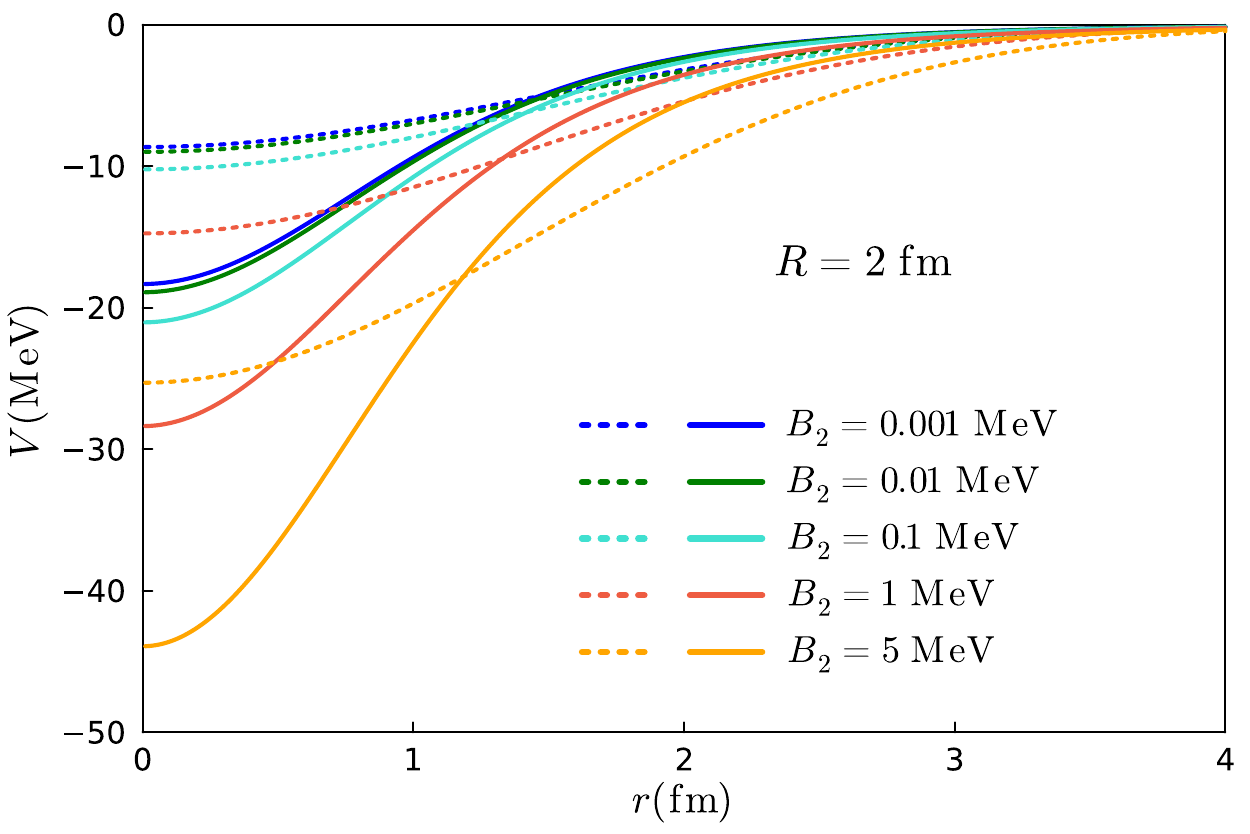}}
  \caption{Two-body potentials for different two-body binding energies. Note that the scales of the axes in these six subfigures are different. The dashed and solid lines are for the Gaussian and tripole potentials.\label{potential}}
\end{figure*}

To estimate the uncertainties of the binding energies of the triple-$J/\psi$ system arising from these parameters, we set the two-body subsystem binding energy $B_2$ (hereinafter abbreviated as two-body binding energy $B_2$) and the cutoff $R$ as follows: $B_2$ ranges from 0.001 to 5 MeV and $R$ ranges from  0.15 to 2 fm. 
The lower and upper bounds of the parameter range are somewhat away from the typical value $R \sim 0.5$ fm.  The reason that we set this wide range is to see the parameter dependence of our study in detail.
The corresponding couplings $C'$ and $C''$ are tabulated in Table~\ref{C-poten}, and the potentials are plotted in Fig.~\ref{potential}. The relation between the coefficient $C'$, $C''$ and $R$ for a fixed binding energy is understood essentially by the well-known formula for the system of a square-well potential of depth $V\ (<0)$ and range $R$, $VR^2 = \pi^2/8\mu$, where $\mu$ is the mass of the particle.  This condition is understood to hold for a bound state of fixed (zero) energy, implying that the potential depth $V$ increases as $1/R^2$,
which, as we observe in Table~\ref{C-poten}, holds well for shallow bound states with short-range interactions.

\begin{table*}[!htbp]
    \setlength{\tabcolsep}{7pt}
    \centering
    \caption{Couplings of the two potentials for different cutoffs $R$ (fm) and two-body binding energies $B_{2}$ (MeV). The values outside and inside the parentheses are for $C'$ and $C''$. $C'$ is in units of MeV and $C''$ is in units of MeV$\cdot$fm$^6$.}
    \begin{tabular}{ccccccccccccc}
    \hline\hline
     $R$ &  $B_{2}=0.001$  & $B_{2}=0.01$  & $B_{2}=0.1$  &  $B_{2}=1$ &  $B_{2}=5$ \\\hline
        0.15 & $-$1502.62 ($-$0.0361) & $-$1506.97 ($-$0.0362)  & $-$1522.42 ($-$0.0366)  & $-$1571.90 ($-$0.0376)  &  $-$1663.67 ($-$0.0394) \\
         0.3 & $-$376.35 ($-$0.579) & $-$378.53  ($-$0.582)  & $-$386.30 ($-$0.592)  & $-$411.48 ($-$0.625)  &  $-$459.33 ($-$0.686) \\
         0.4 & $-$211.96 ($-$1.83) & $-$213.60 ($-$1.84)  & $-$219.44 ($-$1.89)  & $-$238.55 ($-$2.03)  &  $-$275.42 ($-$2.29) \\
         0.5 & $-$135.79  ($-$4.48) & $-$137.13 ($-$4.52)  & $-$141.83 ($-$4.65)  & $-$157.29 ($-$5.07)  &  $-$187.57 ($-$5.87) \\
         0.6 & $-$94.41 ($-$9.30) & $-$95.53 ($-$9.39)  & $-$99.46 ($-$9.72)  & $-$112.49 ($-$10.78)  &  $-$138.38 ($-$12.80) \\
         1 & $-$34.15 ($-$72.05) & $-$34.82 ($-$73.24)  & $-$37.22 ($-$77.50)  & $-$45.39 ($-$91.48)  &  $-$62.51 ($-$118.85) \\
         2 & $-$8.65 ($-$1173.12) & $-$8.99 ($-$1210.50)  & $-$10.23 ($-$1346.57)  & $-$14.75 ($-$1815.69)  &  $-$25.30 ($-$2811.24) \\
    \hline\hline
    \end{tabular}
    \label{C-poten}
\end{table*}

Now, let us turn to the three-$J/\psi$ systems. 
The remarkable Efimov effect, first discovered by Vitaly Efimov in 1970~\cite{Efimov:1970zz}, has been widely studied in nuclear, atomic, and hadronic physics~\cite{Platter:2009gz, Adhikari:1982zz, Lim:1977zz, Valderrama:2018sap, Wu:2020rdg, Ortega:2024ecy}. The likely presence of such an effect in the triple-$J/\psi$ system is worth investigating. A comprehensive review of the Efimov effect can be found in Refs.~\cite{Naidon:2016dpf, Endo:2024cbz, Hammer:2010kp}. Here is a brief introduction. When two particles can form a bound, virtual or resonant state near the threshold through a short-range attractive interaction, an induced long-range three-body attraction emerges, giving rise to an infinite family of three-body bound states, usually called Efimov states or Efimov trimers, with an invariant discrete scale, meaning that the properties of the Efimov trimers are related to each other by a universal scaling factor $\lambda_0=e^{\pi/|s_{0}|}\approx 22.7$, where $s_{0}$ is the solution of the transcendental equation $-s_{0}\mathrm{cosh}(s_{0}\frac{\pi}{2})+\frac{8}{\sqrt{3}}\mathrm{sinh}(s_{0}\frac{\pi}{6})=0$~\cite{Naidon:2016dpf, Endo:2024cbz, Hammer:2010kp}. When the scattering length $a_{sc}\to\pm\infty$ (unitary limit), the binding energies of the Efimov trimers  satisfy the following relation~\cite{Ortega:2024ecy}:
\begin{equation}
    \frac{\bigtriangleup E^{n+1}}{\bigtriangleup E^{n}} \to \frac{1}{\lambda_0^2} \approx 1.94\times 10^{-3}
\end{equation}
where $\bigtriangleup E^{n}=B^{n}_{3}-B_{2}$, $B^{n}_{3}$ are the binding energies of the three-body system, and $B_{2}$ is the binding energy of the two-body subsystem. Note that the binding energies in this work are all defined with respect to the full-dissociation threshold. In the case of large but finite scattering lengths, the number of Efimov states may be limited to a few or even absent altogether. How can we determine whether these three-body bound states qualify as Efimov states based on their energy spectrum? According to Ref.~\cite{Naidon:2016dpf}, a single three-body bound ground state is insufficient to classify it as an Efimov state. Instead, there should be at least two three-body bound states following the universal scaling factor.

\section{Results and Discussions}\label{re-dis}

\begin{table*}[!htbp]
    \setlength{\tabcolsep}{10pt}
    \centering
    \caption{Binding energies of the triple-$J/\psi$ system for various cutoffs $R$ and two-body binding energies $B_{2}$. The values outside and inside the parentheses represent $V(r)=C' e^{-(r/R)^2}$ and $V'(r)=C'' \frac{1}{(r^2+R^2)^3}$. Binding energies are in units of MeV, and cutoffs $R$ are in units of fm.}
    \begin{tabular}{ccccccccccccc}
    \hline\hline
     $R$ &  $B_{3}(B_{2}=0.001)$  & $B_{3}(B_{2}=0.01)$  & $B_{3}(B_{2}=0.1)$  &  $B_{3}(B_{2}=1)$ &  $B_{3}(B_{2}=5)$ \\\hline
    \multirow{2}{*}{0.15} &134.97, 0.298, 0.00347 & 137.68, 0.37, 0.0128 & 147.45, 0.70& 180.24, 2.27 & 246.66, 7.18 \\
         & (188.05, 0.446, 0.00419) & (191.55, 0.54, 0.0142)  & (204.12, 0.94, 0.101) &  (246.05, 2.77) & (329.97, 8.26)  \\
    \multirow{2}{*}{0.3} & 34.17, 0.087, 0.00220 &  35.54, 0.13, 0.0105 & 40.57, 0.33 & 58.32, 1.51 & 97.47, 5.79  \\
        & (47.57, 0.127, 0.00248) & (49.34, 0.18, 0.0110) & (55.78, 0.42) & (78.29, 1.75) &  (126.96, 6.41) \\
    \multirow{2}{*}{0.4} & 19.38, 0.0534, 0.00191 & 20.42, 0.087, 0.0101 &24.25, 0.26 &  38.24, 1.34 &  70.56, 5.51  \\
       & (26.97, 0.0772, 0.00212) &  (28.30, 0.12, 0.0104) & (33.21, 0.32) &  (50.86, 1.53) &  (90.65, 6.02)   \\
    \multirow{2}{*}{0.5} & 12.51, 0.0373, 0.00176 & 13.34, 0.066, 0.0100 & 16.47, 0.22 &  28.19, 1.25 & 56.40, 5.37 \\
       & (17.39, 0.0533, 0.00192) &  (18.47, 0.088, 0.0101)  &  (22.46, 0.27) & (37.18, 1.40) &  (71.62, 5.82)  \\
    \multirow{2}{*}{0.6} & 8.76, 0.0282, 0.00167  & 9.46, 0.053 & 12.11, 0.19  & 22.32, 1.19 &  47.79, 5.30  \\
       & (12.17, 0.0398, 0.00180)  & (13.07, 0.070) & (16.45, 0.23) & (29.22, 1.32)  & (60.08, 5.71)\\
    \multirow{2}{*}{1} &3.25, 0.0137, 0.00153 &  3.69, 0.032 &  5.39, 0.15 &  12.58, 1.09 &  32.55, 5.30 \\
       & (4.53, 0.0189, 0.00165) & (5.08, 0.040) & (7.22, 0.17) & (16.07, 1.18)  & (39.66, 5.65)  \\
    \multirow{2}{*}{2} & 0.89, 0.00637 &  1.12, 0.019 &  2.10, 0.12 &  7.01, 1.05 &  22.88, 6.48  \\
       & (1.28, 0.0105) &  (1.56, 0.026) &  (2.78, 0.14) &  (8.62, 1.13)  & (26.59, 6.36) \\
    \hline\hline
    \end{tabular}
    \label{results}
\end{table*}

\begin{figure}[!htbp]
  \centering
  \begin{overpic}[scale=0.39]{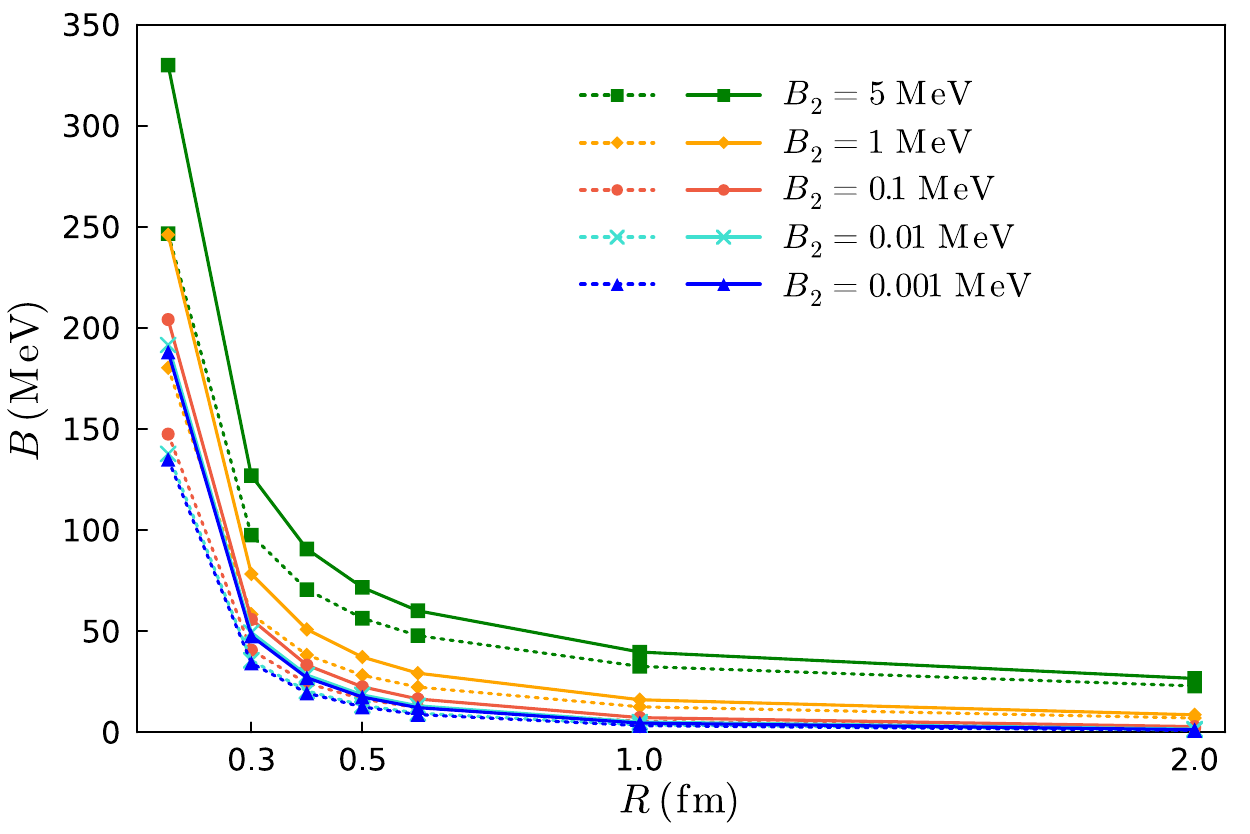}
  \end{overpic}
  \caption{Binding energies of the triple-$J/\psi$ bound ground state as a function of the cutoff $R$ for various two-body binding energies. The dashed and solid lines are for the Gaussian and tripole potentials.}
  \label{Btotal}
\end{figure}

The binding energies of both the ground states and excited states of the triple-$J/\psi$ system for different cutoffs and two-body binding energies are collected in Table~\ref{results}. It is shown that for all the parameter sets, the triple-$J/\psi$ systems are always bound. Furthermore, the triple-$J/\psi$ is more bound for the tripole potential, which may be due to its stronger attraction at shorter ranges than the Gaussian potential as depicted in Fig.~\ref{potential}. 

Fig.~\ref{Btotal} shows the binding energies of the triple-$J/\psi$ system as a function of the cutoff $R$ for various two-body binding energies. 
We observe that the triple-$J/\psi$ binding energies increase as $R$ decreases.  
The reason is as follows.  
Since adding the third $J/\psi$ to the $J/\psi J/\psi$ two-body system makes the distance among $J/\psi$ mesons shorter, each $J/\psi$ pair in the triple-$J/\psi$ system with a smaller $R$ will feel a more attractive interaction, as expected from the behavior  of the potential shown in Fig.~\ref{potential}; as $R$ decreases, the potential depth increases.  
This also explains the different behaviors of the triple-$J/\psi$ system when using the two different potentials.  

When the cutoff $R$ is around the reasonable value of 0.5 fm, even if the two-body subsystem binds very weakly with a binding energy of only 0.001 MeV, the triple-$J/\psi$ still binds with a binding energy of about 15 MeV, implying that the triple-$J/\psi$ bound state is very likely to exist.

In Table~\ref{ev}, we give the expectation values of the Hamiltonian (potential and kinetic energies) and root-mean-square (rms) radii between two $J/\psi$'s of the triple-$J/\psi$ system for different cutoffs $R$ and two-body binding energies $B_{2}$. 
As the cutoff $R$ decreases with fixed two-body binding energy or the two-body binding energy $B_2$ increases with fixed cutoff, the triple-$J/\psi$ system becomes more compact. In addition, the tripole potential gives rise to a more compact triple-$J/\psi$ bound state, which is understood due to its effectively shorter range than the Gaussian one as shown in Fig.~\ref{potential}. 
The results also show that the triple-$J/\psi$ bound states have an equilateral triangle shape. The reason is that we only consider the $S$-wave case and make the symmetrization in exchanging any two $J/\psi$ mesons. 
The size of the triple-$J/\psi$ ground bound state is about 1 fm for a cutoff of $R=$ 0.5 fm, consistent with the typical size of a hadronic molecule given the small size of the $J/\psi$ meson.

\begin{figure}[!htbp]
  \centering
  \begin{overpic}[scale=0.39]{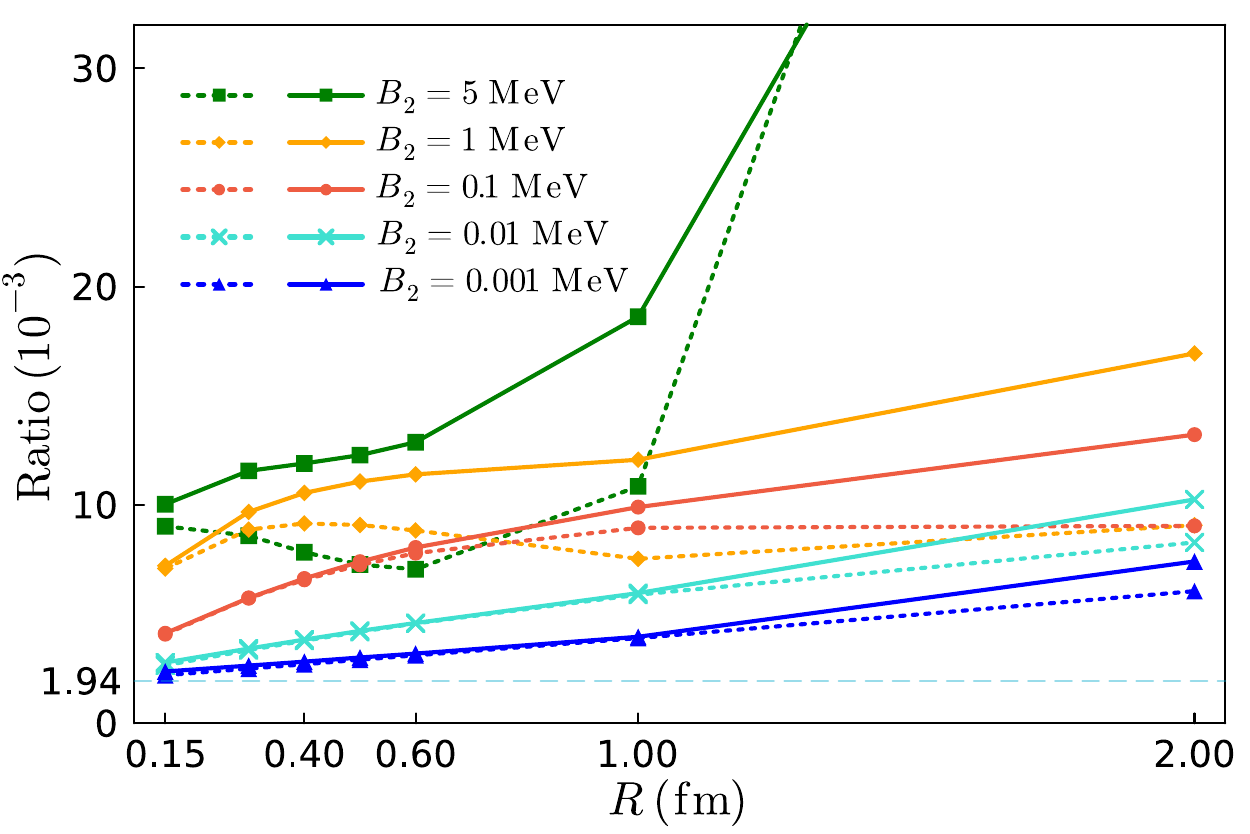}
  \end{overpic}
  \caption{The ratio $\bigtriangleup E^{2}/\bigtriangleup E^{1}$ as a functions of the cutoff $R$ for various two-body binding energies. The dashed and solid lines are for the Gaussian and tripole potentials.}
  \label{ratio}
\end{figure}

To see whether the Efimov effect is realized in the triple-$J/\psi$ system, 
at least two bound states should exist.  This is the case for all the parameter sets that we have chosen in the present study.  As the cutoff $R$ decreases with fixed two-body binding energy or the two-body binding energy $B_2$ decreases with fixed cutoff, a third bound state appears. 
In addition, the tripole potential is more likely to generate a third bound state. 
For instance, for $B_{2}=0.1$ MeV and $R=0.15$ fm, the Gaussian potential gives rise to only two bound states, whereas the tripole potential can give rise to three bound states. As already explained, the reason is that the the triple-$J/\psi$ system with the tripole potential has a larger binding energy.

To confirm whether the obtained two or three bound states are efimov-like states, the energy ratios $\bigtriangleup E^{2}/\bigtriangleup E^{1}$ as a function of various cutoffs, two-body binding energies, and potentials are plotted in Fig.~\ref{ratio}. 
As the cutoff $R$ decreases with a small two-body binding energy, the ratios gradually approach the universal scaling factor for the binding energy $1.94\times 10^{-3}$. This trend is easily understood because a decreasing cutoff indicates a shorter effective interaction range, and a decreasing two-body binding energy means a larger scattering length.
As the cutoff $R$ and two-body binding energy decrease, the differences between the ratios for the two different potentials become smaller, with the ratio for the Gaussian potential being smaller than that for the tripole potential.
For the Gaussian potential, the trends for the two-body binding energies of 1 and 5 MeV have different dips. This difference is because as the two-body binding energy increases, the Efimov effect gradually disappears.

\begin{table*}[!htbp]
    \setlength{\tabcolsep}{10pt}
    \centering
    \caption{Binding energies of the triple-$\Upsilon (1S)$ system for different cutoffs $R$ and two-body binding energies $B_{2}'$. The values outside and inside the parentheses represent $V_{\Upsilon}(r)=C_{\Upsilon}' e^{-(r/R)^2}$ and $V_{\Upsilon}'(r)=C_{\Upsilon}'' \frac{1}{(r^2+R^2)^3}$. The binding energies are in units of MeV, and the cutoffs $R$ are in units of fm.}
    \begin{tabular}{ccccccccccccc}
    \hline\hline
     $R$ &  $B_{3}(B_{2}'=0.001)$  & $B_{3}(B_{2}'=0.01)$  & $B_{3}(B_{2}'=0.1)$  &  $B_{3}(B_{2}'=1)$ &  $B_{3}(B_{2}'=5)$ \\\hline
    \multirow{2}{*}{0.15} & 44.44, 0.1046, 0.00184 & 46.16, 0.1566, 0.0107& 51.84, 0.38 & 71.75, 1.61 & 114.82, 5.96 \\
         & (61.89, 0.1550, 0.00213) & (64.11, 0.2191, 0.0113)  & (71.15, 0.47) &  (96.38, 1.86) & (149.97, 6.59)  \\
    \multirow{2}{*}{0.3} & 11.32, 0.0322, 0.00129 &  12.19, 0.0621  & 15.18, 0.21 & 26.49, 1.23 & 53.94, 5.35  \\
        & (15.74, 0.046, 0.00141) & (16.86, 0.0825) & (20.69, 0.26) & (34.87, 1.38) &  (68.31, 5.78) \\
    \multirow{2}{*}{0.4} & 6.44, 0.0205, 0.00118 & 7.11, 0.0449 & 9.42, 0.18 & 18.57, 1.15 &  42.09, 5.27  \\
       & (8.96, 0.0290, 0.00126) &  (9.81, 0.0583) & (12.75, 0.21) &  (24.14, 1.27) &  (52.43, 5.65)   \\
    \multirow{2}{*}{0.5} & 4.18, 0.0148, 0.00112 & 4.72, 0.0359 & 6.61, 0.16 &  14.46, 1.11 & 35.63, 5.27 \\
       & (5.80, 0.0205, 0.00118) &  (6.48, 0.0456)  &  (8.90, 0.18) & (18.60, 0.21) &  (43.78, 5.63)  \\
    \multirow{2}{*}{0.6} & 2.93, 0.0114, 0.00109  & 3.39, 0.0303 & 5.02, 0.14  & 12.00, 1.08 & 31.59, 5.32  \\
       & (4.07, 0.0157, 0.00113)  & (4.65, 0.038) & (6.71, 0.17) & (15.29, 1.17)  & (38.37, 5.66)\\
    \multirow{2}{*}{1} & 1.11, 0.0060  &  1.39, 0.0204  &  2.48, 0.12 &  7.72, 1.05  &  24.18, 6.06  \\
       & (1.53, 0.0079) & (1.89, 0.0246) & (3.25, 0.14) & (9.56, 1.13)  & (28.37, 6.12)  \\
    \multirow{2}{*}{2} &  0.31, 0.0030  &  0.46, 0.0141  &  1.14, 0.11  &  5.09, 1.16  &   19.24, 8.88   \\
       & (0.43, 0.0038) &  (0.62, 0.0162) &  (1.45, 0.12) &  (6.01, 1.19)  & (21.51, 7.88) \\
    \hline\hline
    \end{tabular}
    \label{results-b}
\end{table*}

It is interesting to extend the triple-$J/\psi$ system to the triple-$\Upsilon (1S)$ system. 
Following the same strategy, i.e., we determine the strength of the potential by reproducing the binding energy of the $\Upsilon (1S)\Upsilon (1S)$ bound state, assuming that two $\Upsilon (1S)$ can form a loosely bound state with a binding energy $B_{2}'$ in the range 0.001 to 5 MeV. For a given two-body binding energy, the potential between two $\Upsilon (1S)$ is smaller than that between two $J/\psi$. This is intuitively clear. As the particle mass increases, the kinematic energy decreases, and consequently, only a smaller attractive interaction is enough to form a bound state with the same binding energy.
The binding energies of both the ground and excited states of the triple-$\Upsilon (1S)$ system are tabulated in Table~\ref{results-b}.  It can be seen that the binding energies of the triple-$\Upsilon (1S)$ system are consistently smaller than those of the triple-$J/\psi$ system for each parameter set. Furthermore, the triple-$\Upsilon (1S)$ system has the same trend for the binding energies with respect to the parameters as the triple-$J/\psi$ system. Moreover, the appearance of the third Efimov state is more difficult in the triple-$\Upsilon (1S)$ system, requiring a shorter cutoff and a smaller two-body binding energy. Considering that the radius of the $\Upsilon (1S)$ meson is only 0.20 fm~\cite{Eichten:1979ms}, a reasonable value for the cutoff $R$ is about 0.2 fm.
In such a case, the binding energies and rms radii of the triple-$\Upsilon (1S)$ bound ground states are 25 $\sim$ 106 MeV and 0.42 $\sim$ 0.25 fm, respectively. It is worth noting that the binding energy of the di-$\Upsilon (1S)$ is predicted to be 300 MeV~\cite{Nefediev:2021pww}, indicating a more bound three-body system. Therefore, a triple-$\Upsilon (1S)$ molecular state is likely to exist.

\begin{table*}[!htbp]
    \centering
    \caption{Expectation values of the Hamiltonian (potential and kinetic energies) and rms radii of the triple-$J/\psi$ system for different cutoffs $R$ and two-body binding energies $B_{2}$. The values outside and inside the parentheses represent $V(r)=C' e^{-(r/R)^2}$ and $V'(r)=C'' \frac{1}{(r^2+R^2)^3}$. Energies are in units of MeV. Cutoffs $R$ and radii are in units of fm.}
    \scalebox{0.96}{
    \begin{tabular}{ccccccccccccc}
    \hline\hline
     $R$ &  $\left \langle T \right \rangle$  & $\left \langle V  \right \rangle$  & $\left \langle r_{ij} \right \rangle$   \\\hline
    \multicolumn{4}{c}{$B_{2}=$ 0.001} \\\hline
    0.15 & 798.13, 25.15, 2.23 (1204.65, 36.51, 2.87) & $-$933.09, $-$25.45, $-$2.23 ($-$1392.70, $-$36.96, $-$2.87) & 0.31, 5.54, 74.42 (0.26, 4.51, 63.97)  \\
    0.3 & 200.54, 6.66, 0.87 (302.71, 9.59, 1.08) & $-$234.71, $-$6.75, $-$0.87 ($-$350.28, $-$9.72, $-$1.08) & 0.62, 10.58, 120.02 (0.52, 8.67, 102.93)   \\
    0.4 & 113.18, 3.89, 0.62 (170.85, 5.57, 0.75) & $-$132.57, $-$3.94, $-$0.62 ($-$197.82, $-$5.64, $-$0.75) & 0.83, 13.70, 148.59 (0.70, 11.27, 125.90)   \\
    0.5 & 72.68, 2.57, 0.48 (109.72, 3.67, 0.57) & $-$85.19, $-$2.61, $-$0.48 ($-$127.11, $-$3.73, $-$0.57) & 1.04, 16.77, 178.69 (0.87, 13.75, 148.69)  \\
    0.6 & 50.64, 1.85, 0.39 (76.45, 2.63, 0.47) & $-$59.40, $-$1.88, $-$0.39 ($-$88.63, $-$2.67, $-$0.47) & 1.24, 19.49, 211.07 (1.04, 16.13, 172.45)   \\
    1 & 18.47, 0.75, 0.24 (27.92, 1.06, 0.28) & $-$21.73, $-$0.77, $-$0.24 ($-$32.44, $-$1.08, $-$0.28) & 2.05, 29.62, 329.32 (1.72, 24.64, 277.60)  \\
    2 & 4.78, 0.25  (7.36, 0.38) & $-$5.67, $-$0.25 ($-$8.64, $-$0.39) & 3.99, 49.05 (3.31, 38.52)  \\\hline
     \multicolumn{4}{c}{$B_{2}=0.01$} \\\hline
     0.15 & 804.47, 27.49, 4.25 (1214.36, 39.41, 5.17) & $-$942.15, $-$27.87, $-$4.26 ($-$1405.91, $-$39.96, $-$5.18) & 0.31, 5.16, 56.18 (0.26, 4.24, 47.71)  \\
    0.3 & 203.70, 7.81, 2.01 (307.55, 11.02, 2.34) & $-$239.24, $-$7.93, $-$2.02 ($-$356.89, $-$11.20, $-$2.35) & 0.62, 9.35, 115.93 (0.52, 7.78, 85.86)   \\
    0.4 & 115.55, 4.73, 1.54 (174.48, 6.63, 1.76) & $-$135.97, $-$4.82, $-$1.55   ($-$202.78, $-$6.75, $-$1.77) & 0.82, 11.78, 219.35 (0.69, 9.86, 128.76)   \\
    0.5 & 74.57, 3.25, 1.25 (112.62, 4.52, 1.43) & $-$87.91, $-$3.31, $-$1.26 ($-$131.08, $-$4.61, $-$1.44) & 1.02, 14.00, 342.88 (0.85, 11.76, 215.64)  \\
    0.6 & 52.21, 2.41 (78.87, 3.33) & $-$61.67, $-$2.46  ($-$91.94, $-$3.40) & 1.21, 16.04 (1.02, 13.52)   \\
    1 & 19.41, 1.08 (29.36, 1.47) & $-$23.10, $-$1.11  ($-$34.43, $-$1.51) & 1.98, 22.98 (1.66, 19.50)  \\
    2 & 5.24, 0.41 (8.06, 0.58) & $-$6.36, $-$0.43  ($-$9.62, $-$0.60) & 3.75, 35.62 (3.11, 29.25)  \\\hline
     \multicolumn{4}{c}{$B_{2}=0.1$} \\\hline
     0.15 & 826.73, 35.45 (1248.50,  49.38,  14.97) & $-$974.19, $-$36.16 ($-$1452.63, $-$50.32, $-$15.07) & 0.31, 4.25(0.26,  3.57,  61.10)  \\
    0.3 & 214.77, 11.69 (324.57, 15.92) & $-$255.34, $-$12.03 ($-$380.34, $-$-16.34) & 0.59, 7.02 (0.50, 5.96)   \\
    0.4 & 123.82, 7.63 (187.21, 10.29) & $-$148.07, $-$7.89  ($-$220.42, $-$10.61) & 0.78, 8.49 (0.65, 7.25)   \\
    0.5 & 81.16, 5.56 (122.78, 7.44) & $-$97.63, $-$5.78 ($-$145.24, $-$7.71) & 0.96, 9.80 (0.80, 8.37)  \\
    0.6 & 57.69, 4.33 (87.32, 5.77) & $-$69.79, $-$4.53  ($-$103.77, $-$6.00) & 1.13, 10.97 (0.95, 9.37)   \\
    1 & 22.65, 2.25 (34.38, 2.96) & $-$28.04, $-$2.40  ($-$41.60, $-$3.13) & 1.78, 14.93 (1.49, 12.63)  \\
    2 & 6.82, 1.02 (10.49, 1.36) & $-$8.92, $-$1.14  ($-$13.26, $-$1.49) & 3.15, 22.53 (2.62, 18.09)  \\\hline
     \multicolumn{4}{c}{$B_{2}=1$} \\\hline
     0.15 & 896.34, 59.84 (1355.75, 80.25) & $-$1076.58, $-$62.12 ($-$1601.80, $-$83.01) & 0.29, 3.00 (0.24, 2.56)  \\
    0.3 & 249.05, 24.00 (377.71, 31.60) & $-$307.37, $-$25.51 ($-$455.99, $-$33.35) & 0.54, 4.57 (0.45, 3.88)   \\
    0.4 & 149.31, 16.98 (226.87, 22.27) & $-$187.55, $-$18.33  ($-$277.72, $-$23.80) & 0.69, 5.43 (0.58, 4.55)   \\
    0.5 & 101.39, 13.15 (154.36, 17.21) & $-$129.59, $-$14.40 ($-$191.54, $-$18.62) & 0.83, 6.19 (0.69, 5.13)  \\
    0.6 & 74.43, 10.74  (113.53, 14.07) & $-$96.75, $-$11.93  ($-$142.76, $-$15.39) & 0.96, 6.91 (0.80, 5.63)   \\
    1 & 32.48, 6.28 (49.92, 8.32) & $-$45.06, $-$7.37  ($-$65.99, $-$9.50) & 1.41, 9.37 (1.18, 7.17)  \\
    2 & 11.57, 3.27 (18.08, 4.51) & $-$18.58, $-$-4.33  ($-$26.71, $-$5.64) & 2.30, 11.72 (1.89, 8.70)  \\\hline
     \multicolumn{4}{c}{$B_{2}=5$} \\\hline
      0.15 & 1019.49, 104.70 (1547.16, 137.61) & $-$1266.14, $-$111.88 ($-$1877.13, $-$145.87) & 0.26, 2.19 (0.22, 1.85)   \\
      0.3 & 309.02, 47.61 (472.03, 62.42) & $-$406.49, $-$53.41 ($-$598.99, $-$68.83) & 0.47, 3.30 (0.39, 2.66)   \\
    0.4 & 193.70, 35.19 (297.14, 46.42) & $-$264.26, $-$40.71  ($-$387.79, $-$52.44) & 0.58, 3.93 (0.48, 3.06)   \\
    0.5 & 136.52, 28.07 (210.29, 37.34) & $-$192.93, $-$33.44 ($-$281.91, $-$43.16) & 0.69, 4.46 (0.57, 3.36)  \\
    0.6 & 103.44, 23.46 (159.95, 31.52) & $-$151.23, $-$28.76  ($-$220.03, $-$37.23) & 0.78, 4.87 (0.65, 3.59)   \\
    1 & 49.46, 15.04 (77.51, 20.71) & $-$82.01, $-$20.34  ($-$117.18, $-$26.36) & 1.11, 5.12 (0.91, 3.92)  \\
    2 & 19.83, 12.58 (31.79, 14.56) & $-$42.71, $-$19.06  ($-$58.38, $-$20.91) & 1.71, 3.77 (1.38, 3.66)  \\
    \hline\hline
    \end{tabular}}
    \label{ev}
\end{table*}

\section{Summary}\label{Con}
This work has studied the possible bound states and Efimov effect of the triple-$J/\psi$ system with the GEM. Because of the expected short-range interaction between the two $J/\psi$ mesons, a contact-range potential was employed. Considering that the radius of the $J/\psi$ is only 0.47 fm, a reasonable value for the cutoff $R$ was estimated to be around 0.5 fm. We then assumed that two $J/\psi$ mesons can form a shallow bound state with binding energies ranging from 0.001 to 5 MeV. 
In addition, the effects of the potential with two different forms on the binding energies and Efimov effect were studied.

The triple-$J/\psi$ system is always bound with a binding energy of 1 $\sim$ 330 MeV for all the parameter sets studied. Note that when the cutoff $R$ has a value of 0.5 fm, even in the case where the interaction between two $J/\psi$ is so weak that a $J/\psi J/\psi$ bound state has a binding energy of only 0.001 MeV, the triple-$J/\psi$ system still binds with a binding energy of about 15 MeV. Considering that the size of the triple-$J/\psi$ ground state is about 1 fm and the radius of the $J/\psi$ meson is only 0.47 fm, we conclude that a triple-$J/\psi$ molecular state is very likely to exist. For the Efimov effect, the triple-$J/\psi$ system has at least two Efimov states for all the parameter sets studied. A third Efimov state appears when the two-body interaction becomes shorter ranged and weaker.

Finally, 
we extended our investigation to the triple-$\Upsilon (1S)$ system, following the same strategy as in the triple-$J/\psi$ system. To form a corresponding two-body bound state with the same binding energy, the potential between two $\Upsilon (1S)$ is weaker than that between two $J/\psi$ due to the larger reduced mass of the $\Upsilon (1S)\Upsilon (1S)$ system. We obtained a triple-$\Upsilon (1S)$ bound state with the same properties as the triple-$J/\psi$ system but with smaller binding energies and fewer Efimov states. With a reasonable cutoff $R$ 0.2 fm, the binding energy and size of the triple-$\Upsilon (1S)$ system is about 65 MeV and 0.34 fm, respectively, implying the likely existence of the triple-$\Upsilon (1S)$ molecular state.
 
 Based on our results, we strongly advocate for experimental efforts aimed at collecting data below the triple-$J/psi$ threshold. This could provide valuable insights into the nature of the triple-$J/psi$ state and contribute to our understanding of the fully heavy multiquark states.

\section{Acknowledgments}
 This work is partly supported by the National Key R\&D Program of China under Grant No. 2023YFA1606700.
Y.W.P. acknowledges support from the China Scholarship Council scholarship and the Academic Excellence Foundation of BUAA for PhD Students
X.L. is supported by 
the National Natural Science Foundation of China under Grant No. 12335001 and 12247101, the project for top-notch innovative talents of Gansu province, the 111 Project under Grant No. B20063, and the Fundamental Research Funds for the Central Universities. 
A.H. is supported in part by the Grants-in-Aid for Scientific Research [Grant No. 21H04478(A), 24K07050(C)].

\bibliography{sample}

\end{document}